\documentclass[aps,amsfonts,superscriptaddress, nofootinbib, floatfix, 10pt,twocolumn,prd]{revtex4-1} 
\usepackage{graphicx}
\usepackage{amsmath}
\usepackage[latin1]{inputenc}
\usepackage{amsbsy}
\usepackage{color}
\usepackage{epsfig}
\usepackage{graphicx}
\usepackage{amsmath}
\usepackage{amssymb}
\usepackage{bbm}
\usepackage{amsbsy}
\usepackage{slashed}

\usepackage{xcolor}

\usepackage[normalem]{ulem}  

\def\be{\begin{equation}}

\def\ee{\end{equation}}
\def\bea{\begin{eqnarray}}
\def\eea{\end{eqnarray}}

\begin{document}


\title{Magnetic Corrections to $\pi$-$\pi$ Scattering Lengths in the Linear Sigma Model}

\author{M.~Loewe}
\email{mloewe@fis.puc.cl}
\affiliation{Instituto de F\'isica, Pontificia Universidad Cat\'olica de Chile, Casilla 306, San\-tia\-go 22, Chile}
\affiliation{Centre for Theoretical and Mathematical Physics and Department of Physics,
University of Cape Town, Rondebosch 7700, South Africa}
\affiliation{Centro Cient\'ifico Tecnol\'ogico de Valpara\'iso-CCTVAL, Universidad T\'ecnica Federico Santa Mar\'ia, Casilla 110-V, Vapara\'iso, Chile} 
\author{L. Monje}
\email{lnmonje@uc.cl}
\affiliation{Instituto de F\'isica, Pontificia Universidad Cat\'olica de Chile, Casilla 306, San\-tia\-go 22, Chile}
\author{R. Zamora}
\email{rzamorajofre@gmail.com}
\affiliation{Instituto de Ciencias B\'asicas, Universidad Diego Portales, Casilla 298-V, Santiago, Chile} 
\affiliation{Centro de Investigaci\'on y Desarrollo de Ciencias Aeroespaciales (CIDCA), Fuerza A\'erea de Chile, Santiago 8020744, Chile}



\begin{abstract}
In this article we consider the magnetic corrections to $\pi$-$\pi $ scattering lengths in the frame of the linear sigma model. For this we consider all the one loop corrections in the s, t and u channels, associate to the insertion of a Schwinger propagator for charged pions, working in the region of small values of the magnetic field. Our calculation relies on an appropriate expansion for the propagator. It turns out that the leading scattering length,  $l=0$  in the S-channel, increases for an increasing value of the magnetic field, in the isospin $I=2$ case whereas the opposite effect is found for the  $I=0$ case. The isospin symmetry is valid because the insertion of the magnetic field occurs through the absolute value of the electric charges. The channel $I=1$ does not receive any corrections. These results, for the channels $I=0$ and $ I=2$   are opposite with respect to the thermal corrections found previously in the literature.

\end{abstract}

\maketitle


\section{Introduction}

Scattering lengths were introduced long time ago in nuclear physics as an important quantity in order to calculate low energy interactions between nucleons and also in pion-nucleon systems. The scattering lengths of two pion systems are relevant  in order to explore QCD predictions in the low energy sector. They were first measured by Rosellet et al \cite{Rosellet}. New measurements have been reported using the formation of pionium atoms in the DIRAC experiment \cite{Dirac}, establishing for the S-wave $\pi$-$\pi $ scattering lengths a 4\% difference between scattering lengths in the isospin channels $I = 0$ and $I =2$.  Another experimental measurement can be explored in the heavy quarkonium $\pi ^{0}$-$\pi ^{0}$ transitions \cite{quarkonium}. In the past, thermal effects on scattering lengths have been considered by many authors in the literature, invoking effective approaches as the Nambu-Jona-Lasinio model \cite{NJL} or the linear sigma model \cite{Martinez}. A common result, at least qualitatively, is that the projection of the scattering lengths in the isospin $I = 0$ channel grows whereas it diminishes in the $I =2$ channel for an increasing temperature.  

\bigskip
In peripherical heavy ion collisions, huge magnetic fields appear. In fact, the biggest fields existing in nature. The interaction between the produced pions in those collision may be strongly affected by the magnetic field. In this article we analyze, in the frame of the linear sigma model, the influence of the magnetic field on the $\pi  $- $\pi $ scattering lengths. For this purpose we will use the weak  field expansion of the bosonic Schwinger propagator \cite{mexicanos}. We present in detail the different analytical techniques we have used for our calculations.



\section{Linear sigma model and $\pi$-$\pi$ scattering}

The linear sigma model was introduced by Gell-Mann and L\'evy \cite{Gell-Mann} as an effective approach for describing chiral symmetry breaking via explicit and spontaneous mechanism.  In the phase where the chiral symmetry is
broken, the model is given by  

\begin{multline}
\mathcal{L}=\bar{\psi}\left[i\gamma^{\mu}\partial_{\mu}-m_{\psi}-g(s+i\vec{\pi}\cdot\vec{\tau}\gamma_{5})\right]\psi\\
+\frac{1}{2}\left[(\partial\vec{\pi})^2+m_{\pi}^2\vec{\pi}^2\right]+\frac{1}{2}\left[(\partial\sigma)^2+m_{\sigma}^2 s^2\right]\\
-\lambda^2vs(s^2+\vec{\pi}^2)-\frac{\lambda^2}{4}(s^2+\vec{\pi}^2)^2+(\varepsilon c-vm_{\pi}^2)s
\end{multline}

In this expression $v=\langle\sigma\rangle$ is the vacuum expectation value of the
scalar field $\sigma$. The idea is to define a new field $s$ such
that $\sigma = s+v$. Obviously $\langle s\rangle=0$. $\psi$
corresponds to an isospin doublet associated to the nucleons,
$\vec{\pi}$ denotes the pion isotriplet field and $c\sigma$ is the
term that breaks explicitly the $SU(2)\times SU(2)$ chiral
symmetry. $\epsilon$ is a small dimensionless parameter. It is
interesting to remark that all fields in the model have masses
determined by $v$. In fact, the following relations are valid:
$m_{\psi}=gv$, $m_{\pi}^2=\mu^2+\lambda^2v^2$ and
$m_{\sigma}^2=\mu^2+3\lambda^2v^2$. Perturbation theory at the
tree level allows us to identify the pion decay constants as
$f_{\pi}=v$. This model has been considered in the context of
finite temperature by several authors, discussing the thermal
evolution of masses, $f_\pi(T)$, the effective potential, etc
\cite{Loewe,Larsen,Bilic,Petropolus,wagner,kovacs1,kovacs2,kovacs3}.

Since our idea is to use the linear sigma model for calculating
$\pi$-$\pi$ scattering lengths, let us remind briefly the
formalism. A scattering amplitude has the general form \cite{libro1,libro2}

\begin{multline}
T_{\alpha\beta;\delta\gamma}=A(s,t,u)\delta_{\alpha\beta}\delta_{\delta\gamma}+A(t,s,u)\delta_{\alpha\gamma}\delta_{\beta\delta}\\
+A(u,t,s)\delta_{\alpha\delta}\delta_{\beta\gamma}
\label{proyectores}
\end{multline}

\noindent where $\alpha$, $\beta$, $\gamma$, $\delta$ denote
isospin components.

By using appropriate projection operators, it is possible to find
the following isospin dependent scattering amplitudes



\begin{align}
T^{0}&=3A(s,t,u)+A(t,s,u)+A(u,t,s)\label{eq3}\\
T^{1}&=A(t,s,u)-A(u,t,s),\label{eq4}\\
T^{2}&=A(t,s,u)+A(u,t,s),
\label{eq5}
\end{align}

\noindent where $T^I$ denotes a scattering amplitude in a given isospin channel.\\

As it is well known \cite{Collins}, the isospin dependent
scattering amplitude can be expanded in partial waves $T_l^I$.

\begin{equation}
T_{\ell}^{I}(s)=\frac{1}{64\pi}\int_{-1}^{1}d(cos\theta)P_{\ell}(cos\theta)T^{I}(s,t,u).
\end{equation}

Below the inelastic threshold the partial scattering amplitudes
can be parametrized as \cite{Gasser}

\begin{equation}
T_{\ell}^{I}=\left(\frac{s}{s-4m\pi^2}\right)^{\frac{1}{2}}\frac{1}{2i}\left(e^{2i\delta_{\ell}^{I}(s)}-1\right),
\end{equation}

\noindent where $\delta_{\ell}$ is a phase-shift in the $\ell$
channel. The scattering lengths are important parameters in order
to describe low energy interactions. In fact, our last expression
can be expanded according to

\begin{equation}
\Re\left(T_{\ell}^{I}\right)=\left(\frac{p^{2}}{m_{\pi}^{2}}\right)^{\ell}\left(a_{\ell}^{I}+\frac{p^2}{m_{\pi}^{2}}b_{\ell}^{I}+...\right).
\end{equation}

The parameters $a_{\ell}^{I}$ and $b_{\ell}^{I}$ are the
scattering lengths and scattering slopes, respectively. In
general, the scattering lengths obey $|a_{0}|>|a_{1}|>|a_{2}|...$.
If we are only interested in the scattering lengths $a_0^I$, it is
enough to calculate the scattering amplitude $T^I$ in the static
limit, i.e. when $s \to 4m_\pi^2$, $t\to 0$ and $u\to 0$

\begin{equation}
a_{0}^{I}=\frac{1}{32\pi}T^{I}\left(s \to 4m_{\pi}^2,t\to 0, u\to0\right).
\end{equation}

\section{One loop magnetic corrections for $\pi$-$\pi$ scattering lengths}

\begin{figure}
\centering
\includegraphics[scale=1]{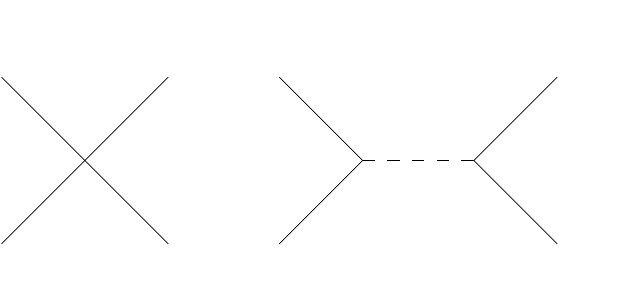}
\caption{Tree level diagrams}
\label{tree}
\end{figure}

The tree level diagrams shown in Fig. \ref{tree}, where the continuum line denotes a
pion, and the dashed line a sigma meson, contribute to the
$\pi$-$\pi$ scattering amplitude. The diagram with a sigma
exchanged meson has
to be considered also in the crossed $t$ and $u$ channels.\\
From these diagrams it is possible to get the results shown in
Table \ref{table1}. The isospin dependent scattering amplitudes at the tree
level have the form

\begin{align}
T^{0}(s,t,u)&=-10\lambda^2-\frac{12\lambda^{4}v^{2}}{s-m_{\sigma}^2}-\frac{4\lambda^{4}v^{2}}{t-m_{\sigma}^2}-\frac{4\lambda^{4}v^{2}}{u-m_{\sigma}^2},\\
T^{1}(s,t,u)&=\frac{4\lambda^{4}v^{2}}{u-m_{\sigma}^2}-\frac{4\lambda^{4}v^{2}}{t-m_{\sigma}^2},\\
T^{2}(s,t,u)&=-4\lambda^2-\frac{4\lambda^{4}v^{2}}{t-m_{\sigma}^2}-\frac{4\lambda^{4}v^{2}}{u-m_{\sigma}^2}.
\end{align}

\begin{table}
\begin{tabular}{|l|p{2.5cm}|p{2.5cm}|p{2.5cm}|}
\hline
 & Experimental Results & Chiral Perturbation Theory & Linear
 Sigma Model
 \\ \hline
$a_0^0$& $0.218\pm 0.02$&$\frac{7m_\pi^2}{32\pi
f_\pi^2}=0.16$&$\frac{10m_\pi^2}{32\pi f_\pi^2}=0.22$\\ \hline
$b_0^0$& $0.25\pm 0.03$&$\frac{m_\pi^2}{4\pi
f_\pi^2}=0.18$&$\frac{49m_\pi^2}{128\pi f_\pi^2}=0.27$\\ \hline
$a_0^2$& $-0.0457\pm 0.0125$&$\frac{-m_\pi^2}{16\pi
f_\pi^2}=-0.044$&$\frac{-m_\pi^2}{16\pi f_\pi^2}=-0.044$\\ \hline
$b_0^2$& $-0.082\pm 0.008$&$\frac{-m_\pi^2}{8\pi
f_\pi^2}=-0.089$&$\frac{-m_\pi^2}{8\pi f_\pi^2}=-0.089$\\ \hline
$a_1^1$& $0.038\pm 0.002$&$\frac{m_\pi^2}{24\pi
f_\pi^2}=0.030$&$\frac{m_\pi^2}{24\pi f_\pi^2}=0.030$\\ \hline
$b_1^1$& $-$&$0$&$\frac{m_\pi^2}{48\pi f_\pi^2}=0.015$\\

\hline

  \hline
\end{tabular}
\caption{Comparison between the experimental values
\cite{peyaud}, first order prediction from chiral perturbation
theory \cite{Weinberg} and our results at the tree level.}\label{table1}
\end{table}

 Note that, the linear sigma model is in a better agreement at tree-level with the
experimental results than first order
chiral perturbation theory.\\


The magnetic corrections to the scattering lengths will be calculated using an appropriate expansion of the Schwinger bosonic propagator wich is given by
\begin{align}
i\Delta (k)&=\int_0^\infty \frac{ds}{\cos(qBs)}e^{ is\left(k_{||}^2-k_\perp^2\frac{\tan(qBs)}{qBs}-m_\pi^2 +i\epsilon \right)}.
\end{align}

\noindent In the above expression, $k_{||}$ and $k_{\perp}$ represent the parallel and perpendicular component of the momentum $k$ with respect to the external magnetic field $B$. In general, as it is well known, this propagator includes a phase factor which, however, does not play any role in our calculation. We proceed by taking the weak field expansion \cite{mexicanos}

\begin{multline}
i\Delta (k)=\frac i {k^2-m_\pi^2+i\epsilon}-\frac {i(qB)^2}{\left(k^2-m_\pi^2+i\epsilon\right)^3}\\-\frac{2i(qB)^2k_\perp^2}{\left(k^2-m_\pi^2+i\epsilon\right)^4}.
\end{multline}

\begin{figure}
\begin{center}
\includegraphics[scale=1]{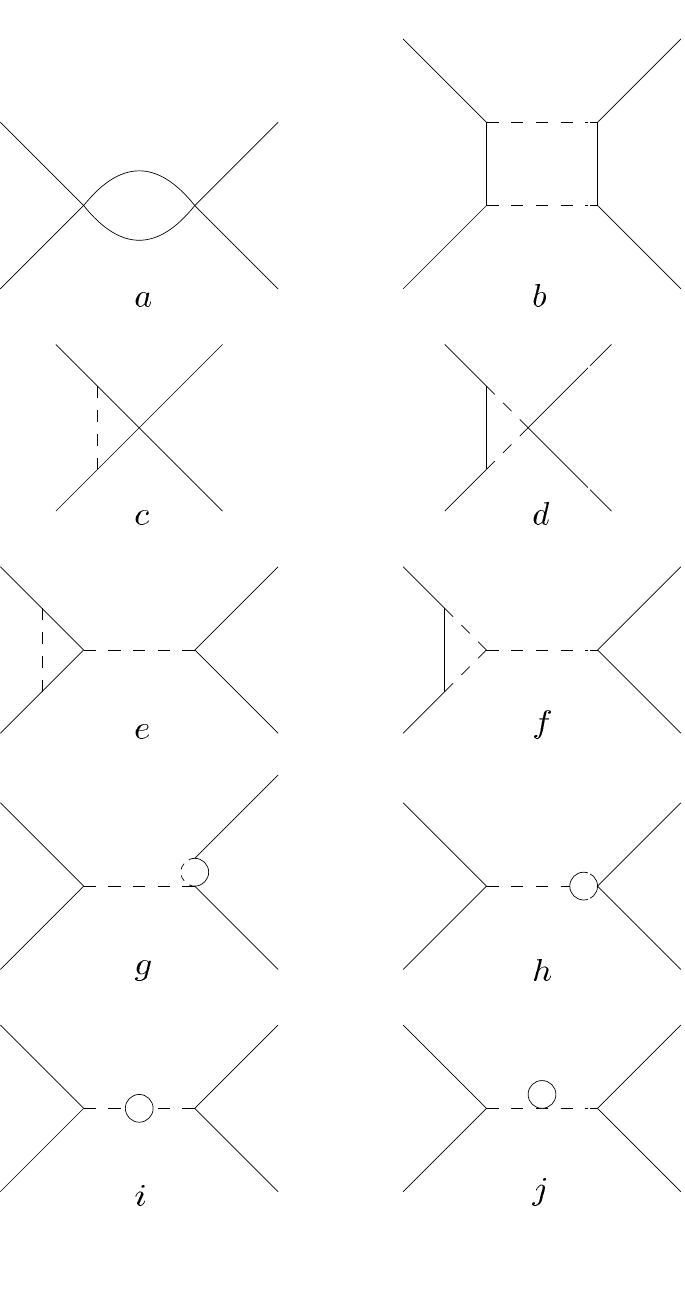}
\caption{Relevant one loop diagrams in the s-channel}
\label{oneloopdiag}
\end{center}
\end{figure}

\noindent Certainly, only the charged pions will receive a magnetic correction. 
There are many diagrams that contribute to the pion-pion
scattering amplitude at the one loop level. For each one of these
diagrams we have to add also the corresponding crossed t and u-channel diagrams. In Fig. \ref{oneloopdiag} we have shown only the s-channel
contribution.

\noindent Explicit expressions will be given only for the diagram (a) and its equivalent diagram in the t-channel.  When
it corresponds, symmetry factors, isospin index contractions, multiplicity factors should be included. We work in the center of mass momentum  $p = (2m_{\pi}, \vec{0})$. For our calculation, since the sigma meson has a much bigger mass than the pions, its propagator will be contracted to a point, the so called high mass limit. A numerical treatment, however, confirms the validity of such approximation in our case. For diagram (a) in Fig. \ref{oneloopdiag} we have

\begin{multline}
i\emph{$M_{a,s}$}=-2\lambda^4 I_{a} \int\frac{d^4
k}{(2\pi)^4} i \Delta(k_{0},\vec{k},m_{\pi})\\\times i \Delta(k_{0}-2m_{\pi},\vec{k},m_{\pi}).
\end{multline}



\begin{figure}
\begin{center}
\includegraphics[scale=0.6]{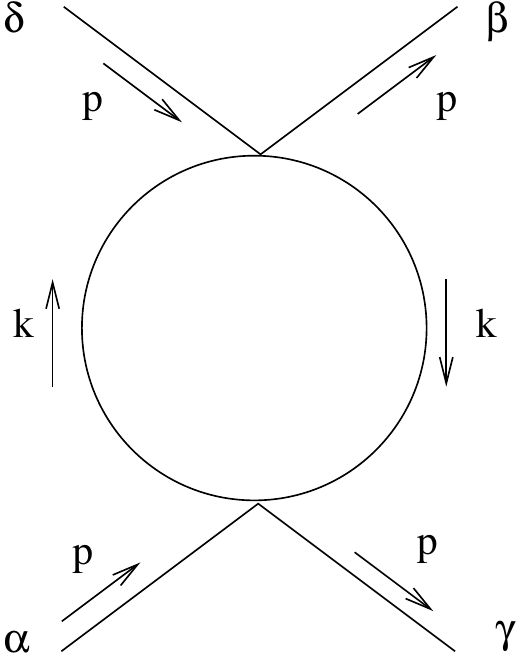}
\caption{Diagram (a) for the corresponding t-channel.}
\label{pescadovertical}
\end{center}
\end{figure}

\noindent The corresponding t-channel diagram in Fig. \ref{pescadovertical}, where no external momentum flows through the loop, is given by
 
\begin{multline}
iM_{a,t}= -4\lambda^4 I_{a}\int \frac{d^4k}{(2\pi)^4} \;\left[i\Delta (k_0,\vec{k},m_\pi)\right]^2, \label{t-channel}
\end{multline}
\noindent  where $I_{a}$ is an isospin term associate to this diagram for the corresponding channel,  that emerges from the contraction of the external pion isospin indexes, which is given by

\begin{align}
I_{a}=(7\delta_{\alpha\beta}\delta_{\gamma\delta}+2\delta_{\alpha\gamma}\delta_{\beta\delta}+2\delta_{\alpha\delta}\delta_{\beta\gamma}),
\label{factorisospin}
\end{align}

\noindent where the greek letters denote isospin indexe. In order to get the scattering lengths in the different isospin channels we have used appropriate projectors contracting them with the amplitudes (See Eq.~(\ref{proyectores})) that emerged from our calculation.

\noindent Notice that for the determination of the scattering lengths we only need the imaginary part of our diagrams. It is natural in this context to choose the $\pi$-$\pi$ center of mass frame of reference for carrying on the calculations in the s-channel. In fact, the  scattering lengths are defined in this frame. Let us consider first the s-channel diagram (a). The idea is to calculate the loop using the weak expansion for the propagators. Then we get expressions that involve free propagators or powers of them. We found quite useful to use the following technique, introduced in \cite{piccinelli}, where a proper time representation for the free propagators is used.  The one loop diagram, at the lowest order, i.e.\ using normal free propagators 

\begin{align}
iL(p)=\int\frac{d^4k}{(2\pi)^4}D(k)D(p-k),
\end{align}
\noindent can be  written in terms of a proper time representation for each propagator
\begin{align}
D(k)=\int_0^\infty ds\; e^{is(k^2-m_\pi^2+i\epsilon)},
\end{align}
as,
\begin{multline}
iL(p)=\int\frac{d^4k}{(2\pi)^4}\int_0^\infty ds_1\;ds_2\;e^{-i(s_1+s_2)m_\pi^2}\\\times e^{is_1(p-k)^2}e^{is_2k^2}.
\end{multline}

\noindent After integrating the Gaussian term in the loop momentum and introducing the following variables
\begin{center}
\begin{tabular}{ccc}
$s_1=s\frac{1-v}2$ & and  & $s_2=s\frac{1+v}2$	
\end{tabular}
\end{center}

\noindent we find that the imaginary part, $\Im{L}$, is given by

\begin{align}
\Im L&=\int_0^1dv\frac1{2\pi i} \int_{-\infty}^{\infty}\;\frac{ds}s\; e^{is\left(\frac14\left(1-v^2\right)p^2-m_\pi^2 \right)}.
\end{align}

\noindent Using the integral representation of the Heaviside function, we obtain
\begin{align}
\Im L&=\int_0^1dv\;\theta\left[\frac14\left(1-v^2\right)p^2-m_\pi^2\right]\nonumber\\
     &=\sqrt{1-\frac{4m_\pi^2}{p^2}}\theta(p^2-4m_\pi^2),
\end{align}

\noindent where $p$ is the total momentum that goes into the loop. If we choose  $p = (2m_{\pi}, \vec{0})$ we see that this contribution vanishes. For higher powers in the denominators, that appear when magnetic field terms are introduced,  we will use the following identity
\cite{Umezawa}

\begin{equation}
\frac{1}{N!}\left(\frac{i\partial}{\partial
\mu ^{2}}\right)^{n}\Delta=\Delta^{n+1}.
\end{equation}

\noindent In this way we found that all the contributions to order $(qB)^{2}$ have the general form

\begin{multline}
L_{\text {general term}} \propto (qB)^2\left(\frac{\partial}{\partial\mu_\pi^2}\right)^2\\\times \int ds_1\;ds_2\; \frac{d^4k}{(2\pi)^4} e^{is_1(k_0^2-\vec{k}^2-m_\pi^2+i\epsilon)}\\ \times e^{is_2((k_0-2m_\pi)^2-\vec{k}^2-\mu_\pi^2+i\epsilon)}.
\end{multline}

\noindent There are several terms of this form where we have to take derivatives with respect to a pion mass parameter, taking then, after the derivation,  all masses as the pion mass. Using the identities shown above, we can see that at the threshold these contributions vanish. So, the s-channel diagram, calculated at the center of mass momentum  $p = (2m_{\pi}, \vec{0})$ does not contribute to the scattering lengths.

\bigskip
\noindent Now we will proceed with the calculation of the equivalent t-channel diagram. At the end, we have also to take into account the u-channel contribution, which is, however essentialy the same as in the t-channel. For this calculation we will invoke some properties of the Hurwitz-$\zeta$ function.  If we consider \eqref{t-channel}, including the coupling, the isospin term and the magnetic field,  integrating the transverse momenta and the proper times we get

\begin{multline}
iM_{a,t}=-\frac{8\lambda^4\pi qB}{(2\pi)^4}\sum_{l=0}^\infty\int dk_0dk_3\;\\\times\frac{(-1)}{\left(qB(2l+1)-k_{||}^2+m_\pi^2-i\epsilon\right)^2}.
\end{multline}
\noindent In the above expression $k _{||}^{2} = k_{0}^{2} - k_{3}^{2}$.  Using the mass derivative, and the Plemelj decomposition $(a \pm i\epsilon)^{-1} = \text{P}(1/a) \mp i\pi\delta(a)$, where P is the Cauchy principal value, we get

\begin{multline}
iM_{a,t}=-\frac{8\lambda^4\pi qB}{(2\pi)^4}\left(\frac\partial{\partial m_\pi^2}\right)\\\times\sum_{l=0}^\infty\int dk_0dk_3\;i\pi\,\delta\left(qB(2l+1)-k_{||}^2+m_\pi^2\right).
\end{multline}
\noindent After some change of variables, the integration in $k_{0}$ gives us 

\begin{multline}
iM_{a,t}=-\frac{8\lambda^4\pi^2}{(2\pi)^4}i\left(\frac\partial{\partial m_\pi^2}\right)\\\times\int dk_3\,\frac{qB}{\sqrt{2qB}} \zeta\left(\frac12,\frac12+\frac{k_3^2+m_\pi^2}{2qB}\right),
\end{multline}

\noindent where we have used the  Hurwitz-$\zeta$ function 

\begin{align}
\zeta(s,q)&=\sum_{n=0}^\infty\frac1{(q+n)^s}.
\end{align}
\noindent We may use the following identity \cite{magnus}

\begin{align}
\zeta(s,a)&=\frac12 a^{-s}+\frac{a^{1-s}}{s-1}+\frac{Z(s,a)}{\Gamma(s)},
\end{align}

\noindent where $Z(s,a)$ in the large-a (Poincar\'e) asymptotic expansion takes the form 
\begin{align}
Z(s,a)\sim\sum_{k=1}^\infty\frac{B_{2k}}{(2k)!}\frac{\Gamma(2k+s-1)}{a^{2k+s-1}}.
\end{align}
\noindent In our case $a = \frac{1}{2} + \frac{1}{2x}$ where $x= \frac{qB}{k_{3}^{2} + m_{\pi }^{2}}$. Notice that a large-a value corresponds to a small magnetic field. Expanding around $x=0$ we get

\begin{multline}
iM_{a,t}=-\frac{8\lambda^4\pi^2qB}{(2\pi)^4\sqrt{2qB}}i\left(\frac\partial{\partial m_\pi^2}\right)\\\times\int dk_3 \left[-\frac{\sqrt2}{\sqrt x}-\frac{x^{3/2}}{12\sqrt2}+O[x]^{7/2}\right].
\end{multline} 

\noindent Keeping only the magnetic contribution $(qB)^{2}$, after integrating in $k_{3}$ finally we find for the t-channel contribution

\begin{align}
iM_{a,t}&=\frac{8\lambda^4\pi^2}{(2\pi)^4}i\left(\frac\partial{\partial m_\pi^2}\right)\left(\frac{(qB)^2}{12m_\pi^2}\right)=-\frac{\lambda^4 i(qB)^2}{24\pi^2m_\pi^4}.
\end{align}

\noindent All the other diagrams reduce to one of the previous cases, once the sigma propagator is cutted, i.e.\ when the approximation $\Delta _{\sigma}(k_0,\vec{k},m_\sigma) \approx -\frac{i}{m_{\sigma}^{2}}$  is used. 

\noindent After taking all the diagrams into account for the one loop magnetic corrections to the $\pi$-$\pi$ scattering amplitudes we get the following amplitude in the s-channel
\begin{multline}
A_B(s,t,u)=\frac{\left(\frac{1}{4 m_\pi^2-m_\sigma^2}\right)^2 \left(\frac{qB}{m_\pi}\right)^2 \left(\lambda ^6 v^2\right)}{4 \pi ^2}\\-\frac{2 \left(\frac{qB}{m_\pi^2}\right)^2 \left(\lambda ^8 v^4\right)}{6 \pi ^2 m_\sigma^4},
\end{multline}
and for the t and u-channels
\begin{widetext}
\begin{multline}
A_B(t,s,u)=A_B(u,t,s)=-\frac{2 \left(\frac{qB}{m_\pi^2}\right)^2 \left(\lambda ^8 v^4\right)}{2 \pi ^2 m_\sigma^4}-\frac{2 \left(\frac{qB}{m_\pi^2}\right)^2 \left(\lambda ^8 v^4\right)}{6 \pi ^2 m_\sigma^4}+\frac{2 \left(\frac{qB}{m_\pi^2}\right)^2 \left(\lambda ^8 v^4\right)}{6 \pi ^2 m_\sigma^4}-\frac{2 \left(\frac{qB}{m_\pi^2}\right)^2 \left(5 \lambda ^6 v^2\right)}{12 \pi ^2 m_\sigma^4}\\-\frac{2 \left(\frac{qB}{m_\pi^2}\right)^2 \left(\lambda ^6 v^2\right)}{12 \pi ^2 m_\sigma^2}-\frac{2 \lambda ^4 \left(\frac{qB}{m_\pi^2}\right)^2}{24 \pi ^2}+\frac{2 \left(\frac{qB}{m_\pi}\right)^2 \left(\lambda ^6 v^2\right)}{4 \pi ^2 m_\sigma^4}.
\end{multline}
\end{widetext}





\section{Results and conclusions}

The magnetic corrections were calculated analytically.
The different parameters in our expressions are renormalized
at $B=0$. The linear sigma model, excluding
the nucleons, has three parameters: $m_{\pi}^2$, $f_{\pi}$ 
and $\lambda^2$. The first two parameters, $m_{\pi}^2$ and $f_{\pi}$, 
are given by experiments and the third one is a free parameter. 
Notice that $f_{\pi}$ is related to the vacuum expectation value $v$. In fact, at the
tree level $f_{\pi}=v$. The three parameters are not independent.
If instead of $f_{\pi}$ we use the vacuum expectation value $v$
and consider a mass of the sigma meson $m_\sigma$=700 MeV, we have $\lambda^2$, $v$= 90 MeV; if
$\lambda^2$=5.6, $v$=120 MeV \cite{Basdevant}. We know, however, that the mass of
the sigma meson is about $m_\sigma$=550 MeV \cite{Aitala}. Therefore,
we need to find new values for $\lambda$ and $v$ associated to the
new lower mass of the sigma meson. We found $\lambda^2$=4.26
and $v$=89 MeV, following the philosophy presented in
\cite{Basdevant}. 
The scattering lengths associated to the isospin channels $I=0$ and $I=2$, including our magnetic corrections,
are given by

\begin{equation}
a_0^0(B)=0.217 + \frac{3 A_B(s,t,u) + 2 A_B(t,s,u)}{32 \pi},
\end{equation}
\begin{equation}
a_0^2(B)=-0.041 + \frac{2 A_B(t,s,u)}{32 \pi}.
\end{equation}

\begin{figure}
\begin{center}
\includegraphics[angle=0,width=91mm]{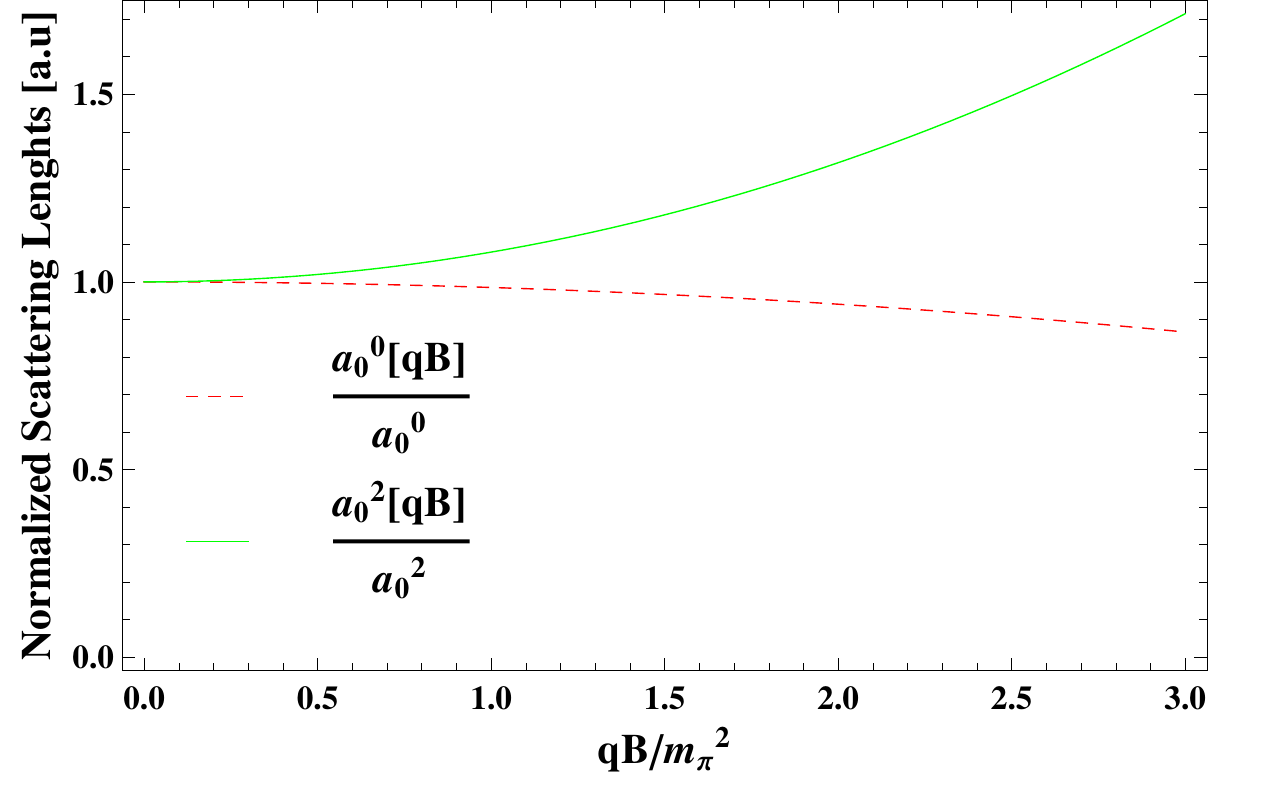}
\caption{(color online) Scaterring lenghts normalized to $B=0$.}
\label{graffinal}
\end{center}
\end{figure}

\noindent The behaviour of the normalized scattering lenghts $\frac{a_0^0(B)}{a_0^0}$ and
$\frac{a_0^2(B)}{a_0^2}$ are shown in Fig. \ref{graffinal} respectively.


\bigskip
The channel $I=2$ corresponds to the most symmetric state for a two pion state in the isospin space. The fact that the scattering length in this channel increases, due to magnetic effects, shows that the interaction between pions become more intense. This, in turn, can be associated to a proximity effect between the pions. In a different context we have found recently a similar effects when computing the correlation distance for quarks in  the quark-gluon plasma. This magnitude increases as function of an increasing external magnetic field \cite{ayala1} being the effect of temperature  exactly the opposite. A similar result was found, this time in the context of QCD sum rules, where the effects of the magnetic field is to increase the continuum threshold \cite{ayala2} whereas temperature induces the opposite effect. We see that the results found in this article are consistent with this general picture.

\section*{ACKNOWLEDGMENTS}


M. Loewe acknowledges support from FONDECYT (Chile) under grants No. 1170107, No. 1150471, No. 1150847 and ConicytPIA/BASAL (Chile) grant No. FB0821, L. Monje acknowledges support from FONDECYT (Chile) under grant No. 1170107 and R. Zamora would like to thank support from CONICYT FONDECYT Iniciaci\'on under grant No. 11160234.



\end{document}